\documentclass[fleqn,usenatbib]{mnras}

\usepackage{newtxtext,newtxmath}
\usepackage[T1]{fontenc}
\DeclareRobustCommand{\VAN}[3]{#2}
\let\VANthebibliography\thebibliography
\def\thebibliography{\DeclareRobustCommand{\VAN}[3]{##3}\VANthebibliography}
\usepackage{graphicx}	% Including figure files
\usepackage{amsmath}	% Advanced maths commands

\title[Puffy vs Failed UDGs]{
Why do some Ultra Diffuse Galaxies have Rich Globular Cluster Systems?}

\author[D. A. Forbes et al.]{Duncan A. Forbes, $^{1}$\thanks{E-mail: dforbes@swin.edu.au}
Maria Luisa Buzzo$^{1,2}$, 
Anna Ferre-Mateu$^{1,3,4}$, Aaron J. Romanowsky$^{5,6}$, 
\newauthor
Jonah Gannon$^{1,2}$, 
Jean P. Brodie$^{1,6}$ and  
Michelle L. M. Collins$^{7}$
\\
$^{1}$ Centre for Astrophysics \& Supercomputing, Swinburne University, Hawthorn, VIC 3122, Australia\\
$^{2}$ ARC Centre of Excellence for All Sky Astrophysics in 3 Dimensions (ASTRO 3D), Australia\\
$^{3}$ Instituto Astrofisica de Canarias, Av. Via Lactea s/n, E38205 La Laguna, Spain\\
$^{4}$ Departamento de Astrofisica, Universidad de La Laguna, E-38200, La Laguna, Tenerife, Spain\\
$^{5}$ Department of Physics \& Astronomy, San José State University, One Washington Square, San Jose, CA 95192, USA\\
$^{6}$ Department of Astronomy \& Astrophysics, University of California Santa Cruz, 1156 High Street, Santa Cruz, CA 95064, USA\\
${^7}$ School of Maths and Physics, University of Surrey, Guildford, UK GU2 7XH
}

\date{Accepted XXX. Received YYY; in original form ZZZ}

\pubyear{2023}

\begin{document}
\label{firstpage}
\pagerange{\pageref{firstpage}--\pageref{lastpage}}
\maketitle

\begin{abstract}
Some ultra diffuse galaxies (UDGs) reveal many more globular clusters (GCs) than classical dwarf galaxies of the same stellar mass.   
These UDGs,  with a mass in their GC system (M$_{GC}$) 
approaching 10\% of 
their host galaxy stellar mass 
(M$_{\ast}$), are also inferred to have high halo mass to stellar mass ratios (M$_{halo}$/M$_{\ast}$). They  have been dubbed  Failed Galaxies. It is unknown what role high GC formation efficiencies and/or low destruction rates play in determining the high 
M$_{GC}$/M$_{\ast}$ ratios 
of some UDGs. Here we present a simple model, which is informed  by recent JWST observations of lensed galaxies and by a simulation in the literature of GC mass loss and tidal disruption  in dwarf galaxies. 
With this simple model, we aim to constrain the effects of GC efficiency/destruction on the observed GC richness of UDGs and their variation with  the integrated 
stellar populations of UDGs. We assume 
no ongoing star formation (i.e. quenching at early times) and that the disrupted GCs contribute their stars to 
those of the host galaxy.  
We find that UDGs, with high 
M$_{GC}$/M$_{\ast}$ ratios today, are
most likely the result of very high GC formation efficiencies combined
with modest rates of GC destruction. 
The current data
loosely follow the model that 
ranges from the mean stellar population of classical dwarfs to that of metal-poor GCs as M$_{GC}$/M$_{\ast}$ increases.
As more data becomes available for UDGs, our simple model can be refined and tested further.

\end{abstract}


\section{Introduction} \label{sec:intro}

A subset of low surface brightness galaxies with large effective radii (R$_e$ $>$ 1.5 kpc) and low central surface brightnesses ($\mu_{g,0}$ $>$ 24
mag per sq. arcsec) have recently obtained much attention. 
These ultra diffuse galaxies (UDGs; \citealt{2015ApJ...798L..45V}) are dwarfs 
with L $<$ 10$^9$ L$_{\odot}$ but UDGs are larger than classical dwarfs which have R$_e$ $<$ 1.5 kpc. UDGs 
have several interesting properties:\\

\noindent
$\bullet$
They tend to be featureless, lacking components clearly seen in other galaxies such as spiral arms, bars, bulges etc
\citep{2017ApJ...844L..11V}.
\\
%$\bullet$
%have typical stellar masses of log M$_{\ast}$ $\sim$ 8 and can be regarded as dwarf galaxies \citep{2022MNRAS.517.2231B}.\\
$\bullet$ 
Found in all environments,  with their number  proportional to the mass of the environment, e.g.
\cite{2017A&A...607A..79V};
\cite{2019ApJ...887...92J};
\cite{2022A&A...665A.105L};
\cite{2023MNRAS.519..884K}.\\
$\bullet$
Cluster UDGs tend to be red, gas-poor and passive, while field UDGs are mostly blue, gas-rich and star forming, e.g. 
\cite{2017MNRAS.468.4039R}; 
\cite{2019MNRAS.488.2143P}.\\
$\bullet$
UDGs in high density environments tend to have formed earlier and faster (as indicated by their alpha-element ratios) than those in low density environments, e.g.  
\cite{RuizLara2018}; 
\cite{Gu2018}; 
\cite{2023MNRAS.526.4735F}.\\
$\bullet$
While some individual UDGs show clear signs of interaction, e.g. \cite{2021MNRAS.502.3144G}; 
\cite{2024ApJ...967L..24O},  most UDGs do not 
\citep{2017ApJ...851...27M}.
\\
$\bullet$
They host up to several times more GCs, per unit galaxy starlight, than classical dwarfs of the same mass, e.g.
%\cite{2020ApJ...899...69L};
\cite{2018ApJ...862...82L}; 
\cite{2020MNRAS.492.4874F}.\\
$\bullet$
Cluster UDGs host more GCs per unit starlight on average than UDGs located in lower density environments, e.g.  
\cite{2020ApJ...899...69L};
%\cite{2018ApJ...862...82L};
\cite{2023ApJ...942L...5J}
.\\
$\bullet$ 
Assuming that they follow the same GC number vs halo mass relation for normal galaxies 
(\citealt{2009MNRAS.392L...1S}; %\citealt{2013ApJ...772...82H}; 
\cite{2017ApJ...836...67H};
\citealt{Burkert2020};
\citealt{2024arXiv240807124L}), GC-poor (N$_{GC}$ $<$ 20) UDGs follow the standard stellar mass--halo mass relation, while the GC-rich (N$_{GC}$ $>$ 20) UDGs reside in overly massive halos 
\citep{2024MNRAS.528..608F}\\
$\bullet$
UDGs with low metallicities for their stellar mass are generally found to be GC-rich, e.g.  \cite{2022MNRAS.517.2231B}; Buzzo et al. (2024, MNRAS, submitted).\\
$\bullet$ GC-rich UDGs tend to be of lower surface brightness \citep{2020MNRAS.492.4874F} and 
rounder, perhaps indicating they are more pressure-supported  
\citep{2024MNRAS.529.4914P}.\\
%(Pfeffer et al. 2024, in press).\\

The presence of two types of UDGs was hinted at in the first study of the GC systems of Coma UDGs \citep{2017ApJ...844L..11V}. Recently, a  
statistical analysis of UDGs by \cite{2024MNRAS.529.3210B} and Buzzo et al. (2024, MNRAS, submitted) found that one type is GC-rich, old, metal-poor, round, with short star formation (SF) timescales, typically found in high density environments, and another type that is GC-poor, younger, has metallicities that match the mass-metallicity relation, with longer SF timescales and typically found in low density environments.

Several models for the formation of UDGs have focused on external processes due to their environment, e.g. 
\cite{2015MNRAS.452..937Y}; 
\cite{2019MNRAS.487.5272J}; 
\cite{2021MNRAS.502..398C}; \cite{2021NatAs...5.1255B}. Such models provide a possible mechanism for tidally puffing-up a dwarf galaxy giving it a larger size for its stellar mass. 
Given that some UDGs are located well outside clusters, there must be a formation mechanism that does not depend on environment. Indeed, several studies have focused on internal mechanisms such as high spin 
\citep{2016MNRAS.459L..51A} 
or episodic feedback from supernovae  \citep{2017MNRAS.466L...1D}. These models are often referred to as {\it `Puffy Dwarf'} models since an  initial classical 
dwarf galaxy can be puffed-up in size, and reduced in surface density, thereby matching the definition of a UDG. 

An alternative is the {\it `Failed Galaxy'} scenario 
(\citealt{2015ApJ...798L..45V}; 
\citealt{2016ApJ...822L..31P}; 
\citealt{2022ApJ...927L..28D}; 
\citealt{2024MNRAS.528..608F}). In this scenario a large number of GCs formed within a substantial dark matter halo but for some unknown reason the galaxy quenched before many field stars could form. 
It could be due to quenching at early times by infall and gas removal as the UDG enters a group/cluster 
(\citealt{2016MNRAS.455.2323M}; 
\citealt{2023MNRAS.518.2453D}), 
although supporting evidence is not clearly 
seen in the currently available data for cluster UDGs ({\citealt{2023MNRAS.525L..93F}
\citealt{2022MNRAS.510..946G}). 
This environmental quenching cannot be the mechanism for isolated UDGs \citep{2022MNRAS.517..858J}. 
Alternatively, quenching might be the result of an interaction with a filament or cosmic sheet 
(\citealt{2023MNRAS.520.2692P}), 
isotropic gas accretion at high redshift 
\citep{2023arXiv231016085K}, or some form of self-quenching due to feedback from an intense period of GC formation so that the gas is no longer available to form stars 
(\citealt{2022ApJ...927L..28D}). 
%(Danieli et al. 2022). 

%Given they started as normal dwarfs, it is not clear how puffy dwarf models could reproduce the GC-rich UDGS, which host more GCs per unit starlight than (non-puffy) dwarfs.  
%models do not reproduce the overly massive halos inferred in some GC-rich UDGs \citep{2024MNRAS.528..608F}, nor  
%do they yet predict the GC system properties of UDGs. 
%The latter leading to the formation of cored dark matter profiles. 

The GC systems of UDGs are clearly a key element in understanding the nature and origin of UDGs. GC richness is often described simply in terms of the total number of GCs (N$_{GC}$) that a galaxy hosts. For dwarf galaxies, the {\it absolute} number can be quite low, even zero
\citep{2018MNRAS.481.5592F}, 
but perhaps a better measure is that of the {\it relative} number. 
Compared to classical dwarf galaxies of the same stellar mass, UDGs have, on average, several times more GCs
(\citealt{2020ApJ...899...69L}; \citealt{2020MNRAS.492.4874F}). 
This difference can be quantified by examining the specific frequency of GCs, S$_N$, which normalises for galaxy luminosity, i.e. 
%the number of GCs per unit M$_V$ = --15 of galaxy starlight. \\
S$_N$ = N$_{GC}$ 10$^{0.4 (M_V + 15)}$.

However, to remove the effects of different stellar populations it is better to use stellar mass. Here 
we use the ratio of GC system mass (M$_{GC}$) divided by host galaxy stellar mass (M$_{\ast}$) expressed as a percentage. We note that other literature works refer to this quantity as S$_M$. In order to convert the number of GCs into a GC system mass, we assume a constant mean GC mass of 2 $\times$ 10$^5$ M$_{\odot}$ for {\it all} dwarf galaxies (similar to that for the GCs of the Milky Way; \citealt{2013ApJ...772...82H}; \citealt{2006ApJ...651L..25J} 
). While there is some evidence that the mean mass of a GC may be slightly smaller in lower mass galaxies \citep{2013ApJ...772...82H}, it is not yet clear if this applies to UDGs. Indeed, one of the most complete luminosity (mass) functions for a GC system associated with a UDG is that of NGC5846\_UDG1 \citep{2022ApJ...927L..28D},   
which reveals a peak luminosity at the same value as the Milky Way.

%At this point, it is useful to give an example. Imagine a dwarf galaxy with a V band luminosity L$_V$ = 10$^8$ L$_{\odot}$, and hence M$_V$ = --15.2, with 24 GCs. Its specific frequency is therefore S$_N$ = 20. 
%If the mean GC has (the turnover magnitude of) M$_V$ = --7.5, then the total luminosity of the GC system is 10$^{4.9}$ x 24 = 10$^{6.3}$. The ratio of the GC system luminosity to that of the host galaxy is therefore 2\%. 
%In this work we assume a mean GC mass of 2 $\times$ 10$^5$ M${_\odot}$, so t
%The total GC system mass in this example would be 24 $\times$ 2 $\times$ 10$^{5}$ M$_{\odot}$. 
%(equivalent to a M/L ratio of $\sim$2).  
%Using M/L$_V$ = 2.5 for the host galaxy would give S$_M$ = 1.9\%. Thus S$_M$ of 2\% is roughly equivalent to S$_N$ = 20, S$_M$ = 5\% to S$_N$ = 50 etc. 

\cite{2020MNRAS.492.4874F} showed that Coma cluster UDGs have a wide range of M$_{GC}$/M$_{\ast}$ ratios from 0\% to $\sim$10\% with an average of $\sim$5\%. Other classical dwarf galaxies also span a range of values, but have averages closer to 1\% for classical Coma dwarfs and 0.5\% for Local Volume dwarf galaxies. For comparison, the stellar halo of the Milky Way has a ratio of around 1--2\% 
\citep{2013MmSAI..84...38L}. This indicates that classical dwarfs cannot be simply puffed-up in size (with no change in total stellar mass or GC content) to explain the GC-rich UDGs with the highest 
M$_{GC}$/M$_{\ast}$ ratios (see also 
\citealt{2022MNRAS.511.4633S}). Recent SED fitting work by Buzzo et al. (2024, MNRAS, submitted) divided UDGs into two classes and found class A (associated with puffy dwarfs) to have an average M$_{GC}$/M$_{\ast}$ = 0.8$\pm$ 0.4\% and class B (associated with failed galaxies) with M$_{GC}$/M$_{\ast}$ = 3.9$\pm$0.5\%. This raises questions, such as
why do some UDGs have much more mass in their GC systems GCs per unit stellar mass than classical dwarfs? Is it due to efficient GC formation or relatively inefficient GC destruction?

%Here, in this work we revisit the toy model presented in \cite{2020MNRAS.492.4874F} to describe the differences in stellar populations expected between puffy dwarfs and failed galaxy UDGs. In particular, 
Here, we investigate whether the stellar population properties of UDGs vary in a systematic way with the observed M$_{GC}$/M$_{\ast}$ ratio today. %ranging from the mean stellar population of classical dwarfs to those of metal-poor GCs. 
In other words, we study UDGs that range from GC-poor puffy dwarfs (with expected low 
M$_{GC}$/M$_{\ast}$) to those of GC-rich failed galaxies (with expected high 
M$_{GC}$/M$_{\ast}$). 
%where disrupted metal-poor GCs may dominate their stellar populations. 
In Section 2 we summarise the current observed upper limit for 
M$_{GC}$/M$_{\ast}$ and in Section 3 describe the properties expected for a failed galaxy.  
%, the ratio of GC system mass to galaxy stellar mass in UDGs. 
Section 4 discusses GC formation and destruction processes. In Section 5 we present a simple model to characterise these processes. We describe our stellar population data and compare it to our simple model predictions in Section 6. Section 7 lists a few examples from the literature of possible failed galaxies.  
Finally, we give our summary and conclusions in Section 8.

%, as the mass of the GC system represents an ever greater fraction of the galaxy stellar mass. 

\section{What is the upper limit to 
M$_{GC}$/M$_{\ast}$ ratio 
for UDGs in the local Universe?}

The spectroscopy-based UDG catalogue of \cite{2024MNRAS.531.1856G} lists two galaxies with apparently 
M$_{GC}$/M$_{\ast}$ $>$10\%. 
They are NGVSUDG-20 and VLSB-B, both located in the Virgo cluster. Both have relatively low stellar masses of $\sim$10$^7$ M$_{\odot}$
\citep{2024MNRAS.531.1856G}. 
NGVSUDG-20 has 
11 GC candidates with a large uncertainty of $\pm$8.6. \cite{2023ApJ...951...77T} confirm 6 GCs and derive a very high velocity dispersion (89 km/s).
%suggests that some of them are not bound to NGVSUDG-20. 
\cite{2024MNRAS.528..608F} 
noted the extreme case of the disturbed galaxy VLSB-B, 
%a galaxy of stellar mass %M$_{\ast}$ = 2.2 $\times$ 10$^7$ M$_{\odot}$ and 
with 26 GC candidates (14 of which have been confirmed spectroscopically \citealt{2023ApJ...951...77T}), 
giving it a ratio of 14\% (confirmed) to 23\% (all candidates). However, again the GC velocity dispersion is quite high (45 km/s). In both cases, the velocity dispersion is very high for the galaxy stellar mass (see fig. 7 of 
\citealt{2023ApJ...951...77T}) which we believe hints at some of the GCs being unbound and/or interlopers.

Recently, 
\cite{2023ApJ...954L..39F} have presented a disturbed UDG (UGC 9050-Dw1) with HST imaging of its GC system, deriving a total GC system of 52. Assuming our method with a universal mean GC mass, the ratio is $\sim$30\% (their own method, based on the ratio of the GC system flux to galaxy flux, places it closer to 20\%). 
%The galaxy mean colour overlaps with the colour range found for the GCs.  
UGC 9050-Dw1 may have one of the highest ratios known, but caution is warranted as the study may have underestimated the background contamination level as it made use of a different HST instrument on an offset field in order to estimate this background. At 35 Mpc it will be extremely difficult to confirm any of its GCs with spectra.

Another extreme galaxy worthy of mention is the IKN dwarf. It is nearby (D $\sim$ 3.6 Mpc) with a size and surface brightness similar to 
those of UDGs. This galaxy has 5 GCs 
(\citealt{2010MNRAS.406.1967G}; \citealt{2024arXiv240511749F}) 
giving it 
M$_{GC}$/M$_{\ast}$ of  %ratio of GC system luminosity to galaxy luminosity of 
$\sim$10\%.  
%(and S$_N$ = 120).  
\cite{2010A&A...521A..43L} have suggested IKN experienced substantial galaxy mass loss due to a past interaction. If this interaction 
has removed more stars than GCs then 
M$_{GC}$/M$_{\ast}$may have been artificially  inflated.
It is also difficult to correctly measure its stellar mass due to a nearby bright foreground star. 
%All of the extreme ratios described above should probably be considered with caution. 

Perhaps the most reliable example of a UDG with a high 
M$_{GC}$/M$_{\ast}$ ratio is NGC5846\_UDG1 (also known as MATLAS-2019). 
From 2-orbit HST imaging the number of GCs was determined to be 54 and the GC luminosity function showed a peak consistent with that of the Milky Way's GC luminosity function, i.e. M$_V$ = --7.5 \citep{2022ApJ...927L..28D}. 
They noted that the 34 brightest, centrally located GCs have an estimated contamination rate of only 0.15 GC.
Using our method  for 54 GCs,  and a stellar mass from the database of 
\cite{2024MNRAS.531.1856G}, 
this corresponds to  M$_{GC}$/M$_{\ast}$ of 
9.8\%. We note that 
\cite{2022ApJ...927L..28D} quote a ratio of 13\%; the difference being due to their higher assumed mean GC mass of 2.5 $\times$ 10$^5$ M$_{\odot}$. 
%We remind the reader than S$_M$ does not depend on assumed distance to a galaxy. 

%In the GC models of Chen 2405.18735, S$_M$ peaks at around 5\% for dwarf galaxies of stellar mass $\sim10^8$ M$_{\odot}$. 
%For such galaxies, the GCs are almost exclusively metal-poor. 

To summarise, there are examples of UDGs in the local Universe today with remarkably high 
M$_{GC}$/M$_{\ast}$ ratios, but, 10\%
is probably the upper limit that we can currently use  with high confidence. While some UDGs show clear evidence for a tidal interaction, the bulk of UDGs do not 
\citep{2017ApJ...851...27M}. 

The original GC system mass in a galaxy is expected to be much higher than today due to the disruption of GCs  
(\citealt{2018RSPSA.47470616F}).
 The stars lost from these GCs would now contribute to the stellar mass of the galaxy. The implication of high 
M$_{GC}$/M$_{\ast}$ ratios today is that the mass contained in GCs at formation was much more significant, and that today the integrated stellar populations of the field stars may be similar to those of old, metal-poor GCs. 
For example, in two simulations of the Fornax dwarf galaxy by \cite{2023MNRAS.522.5638C}, 
the GC system was initially up to
5-8$\times$ more numerous than the final number of half a dozen, due to disruption over cosmic time. This is supported by detailed observations of the chemical abundances of the Fornax GCs which place an upper limit on the original mass of the GC system to be 4--5x that of today \citep{2012A&A...544L..14L}. 

\section{High M$_{GC}$/M$_{\ast}$ galaxies as Failed Galaxies}

As mentioned in the Introduction, a `failed galaxy' is a proposed pathway for UDGs that formed in a massive halo with an effective radius similar to that of a giant galaxy, but,  for some reason, formed a stellar mass more similar to that of a dwarf galaxy. Massive halos are associated with rich GC systems for normal galaxies 
\citep{Burkert2020}, thus failed galaxy UDGs are also expected to host large numbers of GCs for their stellar mass, i.e. extremely high 
M$_{GC}$/M$_{\ast}$ ratios. 

The large number of GCs may have played a key role in self-quenching the galaxy  \citep{2024MNRAS.528..608F}. We might expect 
the GC system to have a similar radial extent as that of its host galaxy given that disrupted GCs can make a significant contribution to its field stars. 
%M$_{GC}$/M$_{\odot}$ ratio.
Failed galaxies may have formed GCs very efficiently, perhaps due to high gas pressures 
\citep{2012MNRAS.426.3008K}
which in turn is reflected in rounder dwarf galaxies that are pressure-supported {\citep{2024MNRAS.529.4914P}. 
Indeed, there is some evidence that rounder dwarf galaxies do have higher 
M$_{GC}$/M$_{\ast}$ values \citep{2024MNRAS.529.4914P}.
%Efficient GC formation may be reflected in a short timescale for star formation. 
Failed galaxies may also have stellar populations that resemble metal-poor GCs 
and so lie below the standard dwarf galaxy mass-metallicity relation. 
Observations provide some support for this with UDGs of low metallicity (for their stellar mass) found to be GC-rich, i.e.  N$_{GC}$ $>$ 20 
(\citealt{2022MNRAS.517.2231B} \citealt{2023MNRAS.526.4735F}). 
%They associated these two types with failed galaxies and puffy dwarfs respectively. 
\cite{2024MNRAS.528..608F} 
found that such GC-rich UDGs were most likely hosted in  overly massive halos with cored profiles. Overly massive halos lie off the standard stellar mass -- halo mass relation in the direction of 
galaxies that have assembled late with low halo concentrations and/or been quenched early. \\

\noindent
In summary, failed galaxies might be expected to have:\\
$\bullet$ High M$_{halo}$/M$_{\ast}$ ratio\\
$\bullet$ Rich GC systems\\
%(N$_{GC}$ $\ge$  20)\\
$\bullet$ High M$_{GC}$/M$_{\odot}$ \\
$\bullet$ Field stars with GC-like stellar populations (i.e. low metallicity, old ages and enhanced in alpha elements)\\
$\bullet$ Relatively flat, or even rising, stellar population gradients.\\
%$\bullet$ field stars of low metallicity given the galaxy's stellar mass\\
$\bullet$ GC system with a similar size to the host galaxy (R$_{GC}$/R$_e$ $\sim$ 1)\\
%$\bullet$ initial star formation with a short burst\\
%$\bullet$ high initial ISM gas pressure\\
$\bullet$ A round host galaxy\\
$\bullet$ Pressure-supported stellar kinematics\\
%$\bullet$ late assembly, low halo concentrations and/or early quenching.\\

Here we explore how the stellar populations of UDGs vary with M$_{GC}$/M$_{\ast}$ ratio. 
We note that \cite{2016ApJ...822L..31P} first coined the phrase `pure stellar halo' for the Coma cluster UDG DF17 which they described as a sub-L$^{\ast}$ failed galaxy. It has an    
M$_{GC}$/M$_{\ast}$ ratio of about 2.5\%. %which we use as an arbitrary division between low and high 
%M$_{GC}$/M$_{\ast}$ratios, and by proxy between puffy dwarfs and failed galaxies. 
In the next sections we discuss GC formation and destruction processes and create a simple model for how 
M$_{GC}$/M$_{\ast}$ depends on the interplay of these two processes.

\section{Globular Cluster Formation and Destruction Processes}

The number of GCs, or the mass of the GC system, observed today is a combination of GC formation, accretion (e.g. mergers), removal (e.g. tidal stripping), mass loss and disruption over time. For dwarf galaxies, accretion of fully-formed, ex-situ GCs (and stars) is expected to make only a small contribution, if any, to the ratio of GC mass to galaxy mass (given that GCs and galaxy stars will be accreted in similar proportions). Similarly, the ratio would remain largely unchanged in tidal stripping since the GC system and stars have a similar radial extent in dwarfs  
(\citealt{2017MNRAS.472L.104F}; 
\citealt{2018MNRAS.477.3869H}; 
\citealt{2022MNRAS.511.4633S}), and so would be removed in similar proportions. 

Here we briefly discuss what processes determine GC formation and what processes that control their destruction (and thus how their stars contribute to the field star population of their host galaxy).

The formation efficiency of GCs is strongly related to the overall star cluster formation efficiency (CFE). The CFE is defined as the fraction of bound clusters as a fraction of the total stellar mass formed.
The model of \cite{2012MNRAS.426.3008K} predicted that CFE is primarily driven by the gas surface density in the host galaxy, and good  agreement was shown between their predicted CFE and observations in the local Universe. This formation model is incorporated into the EAGLE cosmological simulation, along with GC destruction, to predict the properties of GC systems 
\citep{2018MNRAS.475.4309P}. 

%Given the model dependence of CFE on gas density/pressure, 
%\cite{2024MNRAS.529.4914P}
%has shown that GC-rich galaxies should be more pressure-supported than GC-poor galaxies.  
%Other secondary parameters include 
%While gas density/pressure is a key parameter in determing the CFE (Kruijssen2012), t
%There is considerable scatter in the observed CFE vs SF rate surface density relation suggests additional parameters at work.
%Anderson et al. (2023) 2308.12363 
%\cite{2024A&A...681A..28A} 
%examined the effects of feedback on CFE in an N-body model of a dwarf galaxy. They found that as additional sources of feedback (e.g. winds and radiation) were added the CFE was reduced. Thus galaxies that initially formed a high fraction of stellar mass in GCs are those that had minimal feedback. 

In the Gnedin models, e.g. \cite{2023MNRAS.522.5638C}, 
GC formation is dependent on the accretion rate and gas mass. Similarly, the EMERGE-based model of 
\cite{2021MNRAS.505.5815V}
also assumes that when the accretion rate passes some threshold, GC formation is triggered. 
\cite{2024arXiv240518735C} noted that low metallicity gas also acts to enhance the fraction of stars formed in bound clusters.  
Therefore GC formation is most efficient at higher redshifts where high gas pressures and accretion rates are combined with low metallicities. 
Any subsequent star formation after this epoch would tend to reduce the GC mass to stellar mass ratio as it will be less efficient at forming GCs relative to field stars. 

%As well as this smooth accretion, they explicitly  highlight the additional contribution to GC formation induced by mergers. In general, GC-rich galaxies are those that experience above average accretion. 

%At the earliest epoch of GC formation, when GC formation was most efficient, it was suggested that the mass in GCs can be as much as half of the total stellar mass \citep{2017MNRAS.469L..63R}.  
%Theoretical models suggested that the CFE in the early Universe could have fractions up to 70\% \citep{2012MNRAS.426.3008K}. 
%Such high ratios have been supported recently by JWST observations of lensed galaxies (which have the advantage of being randomly selected galaxies).

As noted above, the GC content of a galaxy today is not simply a function of its formation efficiency but also the efficiency by which GCs are disrupted. 
The destruction of GCs over time has been described in classic works such as \cite{1988ApJ...335..720A} and 
\cite{1997ApJ...474..223G}. 
Briefly, GCs are subject to internal and external destruction processes.
Internal ones include two-body relaxation and mass loss due to stellar evolution. %These processes tend to preferentially destroy low mass (and low concentration) GCs. 
Mass loss dominates in the early stages of GC formation. 
External processes are due to the tidal field of the host galaxy \citep{2008MNRAS.389L..28G} -- these can include disk and bulge shocking which remove stars when the GC experiences the tidal field of a disk or bulge. These processes all tend to reduce the mass of individual GCs, the total mass of the GC system and ultimately the number of GCs present with shocks being the most efficient process at early times. These processes evolve an initial power-law distribution into the well-known Gaussian GC luminosity function 
\citep{2023MNRAS.522.5638C}. 
Any stars liberated from disrupted GCs will add to the field star population of the host galaxy. 
%In the case of the Milky Way's halo, such stars may have contributed 25\%, or more, to all halo stars \cite{2017MNRAS.465..501S}. 

Dynamical friction is also an external process, which tends to cause in-spiralling of the more massive GCs, and may lead to the formation of a galaxy nucleus from merged GCs (e.g. \citealt{2022A&A...658A.172F}).
In this case, 
%the total mass of the GC system is unchanged, but the 
the number of GCs is reduced. The GC system mass will also be reduced if the nucleus is counted separately. With some nuclei located slightly off centre, they may, or may not, be included in the GC system count.
The importance of dynamical friction in modifying a GC system in a dwarf galaxy depends strongly on the dark matter density profile -- cuspy profiles, with high density inner regions, will tend to accelerate the in-spiralling process, whereas it will be much reduced for lower density cored profiles.  For example, a core profile has been invoked to explain the presence of several old-aged GCs in the Fornax dwarf galaxy, e.g. \cite{2012MNRAS.426..601C}.

Dwarf galaxies have much weaker tidal fields than giant galaxies, and in the case of UDGs, may lack structures such as disks and bulges, so tidal shocks are expected to be much less important in disrupting the GC systems of dwarfs. Using a cosmological simulation with an N-body code, \cite{2024MNRAS.527.2765M} 
recently modelled the evolution of GCs in classical dwarf galaxies including both internal mass loss and disruption from external environmental effects. Starting with a power-law distribution of cluster masses they followed GCs within dwarf galaxies of different total masses. They found that lower mass, and lower density,  galaxies had lower GC disruption rates. 
%so that 25\% of the initial GC system mass remained (75\% destroyed) in their lowest mass galaxy vs only 12\% (88\% destroyed) in the highest mass galaxy. 
%They found lower density (mass within the half mass radius) galaxies had lower GC disruption rates. 
The latter might explain why UDGs, with lower stellar densities than classical dwarfs, have more GCs per unit starlight today  \citep{2020MNRAS.492.4874F}. 
We note that their model galaxies were drawn from a cosmological simulation and so follow the standard stellar mass--halo mass relation by design. 

Finally, we note that the deep HST imaging of  NGC5846\_UDG1 reveals a standard GC luminosity function shape 
{\citep{2022ApJ...927L..28D}. 
%with a peak around M$_V$ --7.5 (M $\sim$ 2 $\times$ 10$^5$ M$_{\odot}$). 
This implies that the destruction processes that have modified an initial power-law mass distribution have  operated in NGC5846\_UDG1 in a similar overall manner to that of other dwarf and giant galaxies.

\section{A Simple Model for GC Formation and Destruction}

Based on the discussion above, we now describe a simple model for the evolution of the GC-to-stellar mass ratio over cosmic time to better understand the high ratios observed for some UDGs today. 
We assume that the initial mass formed in GCs (M$_{GC,i}$) to that in field stars (M$_{\ast,i}$) at high redshift is equivalent to the GC formation efficiency ($c$) i.e.\\

\begin{equation}
    M_{GC,i} / M_{\ast,i} = c
\end{equation}

\noindent
%Guided by simulations and recent observations with JWST, we set c to be 50\%.\\

%    \begin{equation}
%    and we set it to 50\%, 
%    c = 0.5
%    \end{equation}
%M$_{GC,i}$ / M$_{\ast,i}$ = c = 0.5\\

\noindent
We further assume that due to destruction processes, the current final mass in GCs (M$_{GC,f}$) is some fraction of that initially, i.e. \\

\begin{equation}
M_{GC,f} = (1 - d) M_{GC,i}\\
\end{equation}

\noindent
where $d$ is the destruction fraction. 

Assuming that the galaxy is quenched early with no further star formation and no loss/gain of field stars (e.g. due to tidal interactions or accretion), then the current field star mass is simply the original one plus the mass of stars disrupted from GCs, i.e.,\\

\begin{equation}
M_{\ast,f} = M_{\ast,i}  + d~M_{GC,i}\\
\end{equation}

\noindent
The ratio observed today of GC system mass to stellar mass in percentage terms (or S$_M$) is:\\

\begin{equation}
M_{GC} / M_{\ast} = S_M (\%)\\
\end{equation}

\noindent
Replacing the above we have\\

\begin{equation}
    M_{GC}/M_{\ast}    = (M_{GC,i} - d~M_{GC,i}) ~/~ (1/c~M_{GC,i} + d~M_{GC,i})\\
\end{equation}

\noindent
or\\

\begin{equation}
M_{GC}/M_{\ast} = (1 - d) / (1/c + d)\\
\end{equation}

Equation 6 shows that 
M$_{GC}$/M$_{\ast}$ today in this simple model, depends on only the GC formation efficiency ($c$) and the destruction fraction ($d$). For example, in the absence of any destruction ($d$ = 0), 
M$_{GC}$/M$_{\ast}$ today equals M$_{GC}$/M$_{\ast}$ at formation ($c$ $\times$ 100). If no GCs form ($c$ = 0), then 
M$_{GC}$/M$_{\ast}$
= 0.  

Similarly, the mass in disrupted GCs relative to the final stellar mass of the galaxy is\\

\begin{equation}
d~M_{GC,i}/M_{\ast} = d / (1/c + d)
\end{equation}
}

GC formation was most efficient at the earliest epochs when gas surface densities were high. 
In the models of \cite{2019MNRAS.488.5409C}, star cluster formation is some 20$\times$ more efficient at z = 10 compared to today due to high rates of gas accretion at early times. 
In the models of 
\cite{2012MNRAS.426.3008K} 
the CFE in the early Universe has fractions of up to 80\%.  
At formation, a proto-GC may have 
contained 4--5 
\citep{2012A&A...544L..14L} to 10 
\citep{2017MNRAS.469L..63R}
times as much mass as a GC today. 
\cite{2017MNRAS.469L..63R} inferred that the mass in proto-GCs was roughly equal to half of the total stellar mass. 
%The GC formation efficiency ($c$) 
%may even approach unity,   
%implying that the mass of GCs formed is the same as that of field stars. 
If star formation was highly preferenced to be in GCs over that of individual field stars and if proto-GCs were originally significantly more massive  
then the GC formation efficiency ($c$) could in principle exceed unity.

Observational constraints on $c$ shortly after the formation of GCs are now available from JWST observations of lensed galaxies (although only a few examples are available, they have the advantage of being largely randomly selected galaxies). Several recent examples of galaxies at large look-back times are now available.   
For the `Cosmic Grapes' galaxy at z $\sim$ 6, the light in the star forming clumps is some 70\% of the total light \citep{2024arXiv240218543F}. The `Firefly Sparkle' galaxy 
\citep{2024arXiv240208696M}, with a stellar mass of 4 $\times$ 10$^6$ M$_{\odot}$, is at a redshift z = 8.3 corresponding to a look-back time of over 13 billion years. Its ten star clusters have typical ages of $\sim$100 Myr and masses from 2 to 6 $\times$ 10$^5$ M$_{\odot}$. The total mass of these 10 star clusters is 49--57\% of the total stellar mass of the host galaxy.
%Their location and orbits will determine whether they survive over cosmic time or whether they are disrupted, contributing to the stellar component of the Firefly Sparkle.
The `Cosmic Gems' galaxy at z $\sim$ 10.2 formed only 460 Myr after the Big Bang \citep{2024arXiv240103224A}. 
Its half dozen star clusters have individual masses of 1--2.6 $\times$ 10$^6$ M$_{\odot}$, and a combined stellar mass of around 30\% that of the host galaxy. 
%This fraction rises to $\sim$60\% if the flux in the F150W filter is compared. 
The star clusters have compact sizes of 1--2 pc and of high density suggesting they are gravitationally-bound. 
These JWST observations are likely detecting the brightest GCs shortly after their formation. Lower mass clusters are likely present but undetected implying that the current fractions are lower limits to the true value. Another implication from the `Cosmic Gems' galaxy at z = 10.2, is that the original proto-GCs have already undergone significant mass loss so that they start to resemble bona fide GCs in terms of their sizes and masses only $\le$ 500 Myr after formation.
 This early phase of GC formation appears to be associated with a short-lived but intense period of star formation \citep{2024MNRAS.529.3301T}.

To constrain the destruction fraction of GCs we must use theoretical models.  
In their model of dwarf galaxies, \cite{2024MNRAS.527.2765M} found GC system total destruction rates (i.e. including mass loss and tidal disruption) of 70--88\% after 10--12 Gyr. They also found that low density galaxies, like UDGs, have the lowest destruction rates. Based on their findings, we assume variations in $d$ from 0.7 to 0.9 for UDGs. A destruction rate of 1.0 would mean no GCs at the present day, and so 
M$_{GC}$/M$_{\ast}$ = 0. We note for the MW, the contribution of stars from disrupted GCs to the stellar mass of the halo is estimated to be at least 25\% 
\citep{2017MNRAS.465..501S}.  
%which suggests a destruction rate of $d$ $\ge$ 0.75. 

%The most extreme galaxies discussed above (VLSB-B, UGC 9050-Dw1, IKN) have S$_M$ of 13--35\%, although these values may have been lower before an tidal interaction.
%The observational work of \cite{2012A&A...544L..14L} for the Fornax dwarf galaxy indicates M$_{GC,f}$ = 0.20--0.25 x M$_{GC,i}$, corresponding to S$_M$ = 29--27\%. 

To explore how GC formation and destruction affects 
M$_{GC}$/M$_{\ast}$ in our simple model, we show in Fig.
\ref{fig:one} the dependency of 
M$_{GC}$/M$_{\ast}$
for fixed  destruction fractions of $d$ = 0.7, 0.8 and 0.9 and a variable GC formation efficiency $c$. The plot shows that UDGs with low 
M$_{GC}$/M$_{\ast}$ ratios (i.e. puffy dwarfs) may have experienced very low GC formation efficiencies and/or very high destruction fractions. In order to achieve the observed upper limit of 
M$_{GC}$/M$_{\ast}$ $\sim$ 10\% (corresponding to failed galaxies), destruction fractions of $\sim$70\% (80\%) require modest GC formation efficiencies of $\ge$ 40\% (80\%). For 
destruction fractions of 90\% efficiencies over 100\% would be needed to reach 
M$_{GC}$/M$_{\ast}$ $\sim$ 10\%. 
These GC destruction fractions and formation efficiencies are consistent with the dwarf galaxy simulations of 
\cite{2024MNRAS.527.2765M} and recent JWST observations of high redshift lensed systems.
%The plot also shows a line for no destruction of GCs ($d$ = 0), with 
%M$_{GC}$/M$_{\ast}$  today being equal to 
%M$_{GC}$/M$_{\odot}$ ($c$ $\times$ 100) at formation. 

If we hold the destruction fraction constant, then higher GC formation efficiencies are associated with higher 
M$_{GC}$/M$_{\ast}$  ratios. 
However, at a fixed GC formation efficiency, higher 
M$_{GC}$/M$_{\ast}$ ratios  require a {\it decrease} in the destruction fraction. Thus $c$ and $d$ have an inverse relationship with each other in terms of how they affect the 
M$_{GC}$/M$_{\ast}$ ratio. 

%{\it We conclude that GC-rich galaxies,  
%with S$_M$ of $\sim$10\% (like NGC5846\_UDG1), require modest GC destruction rates ($\sim$70\%) and/or very high GC formation efficiencies ($\ge$ 40\%). }

Globular cluster formation and destruction processes might be expected to lead to  differences in the integrated stellar populations of the galaxy field stars as 
M$_{GC}$/M$_{\ast}$  changes.  At a given GC formation efficiency, higher GC destruction fractions would imply that more GCs have been disrupted, contributing to the field stars of the galaxy, giving rise to more GC-like integrated stellar populations and lower ratios since M$_{GC}$ is reduced while M$_{\ast}$ increases. However, for a fixed modest destruction fraction, higher GC formation efficiencies are associated with higher
M$_{GC}$/M$_{\ast}$ ratios and the integrated stellar populations become more GC-like due to the high rate of GCs that have been formed and then disrupted.
We explore this further in the next section.

%The latter case is supported by the UDG with the highest reliable $S_M$ value, i.e. NGC5846\_UDG1 with its stars having similar stellar populations to its GCs \citep{2020A&A...640A.106M}.
%, suggesting that high S$_M$ UDGs have high destruction rates. (
%Although colours of stars and GCs for an individual UDG are becoming available (Janssens24 submitted), more stellar population measures of stars and GCs are needed for UDGs with a range of S$_M$ values. 

\begin{figure}
	\includegraphics[width=0.8\linewidth, angle=-90]{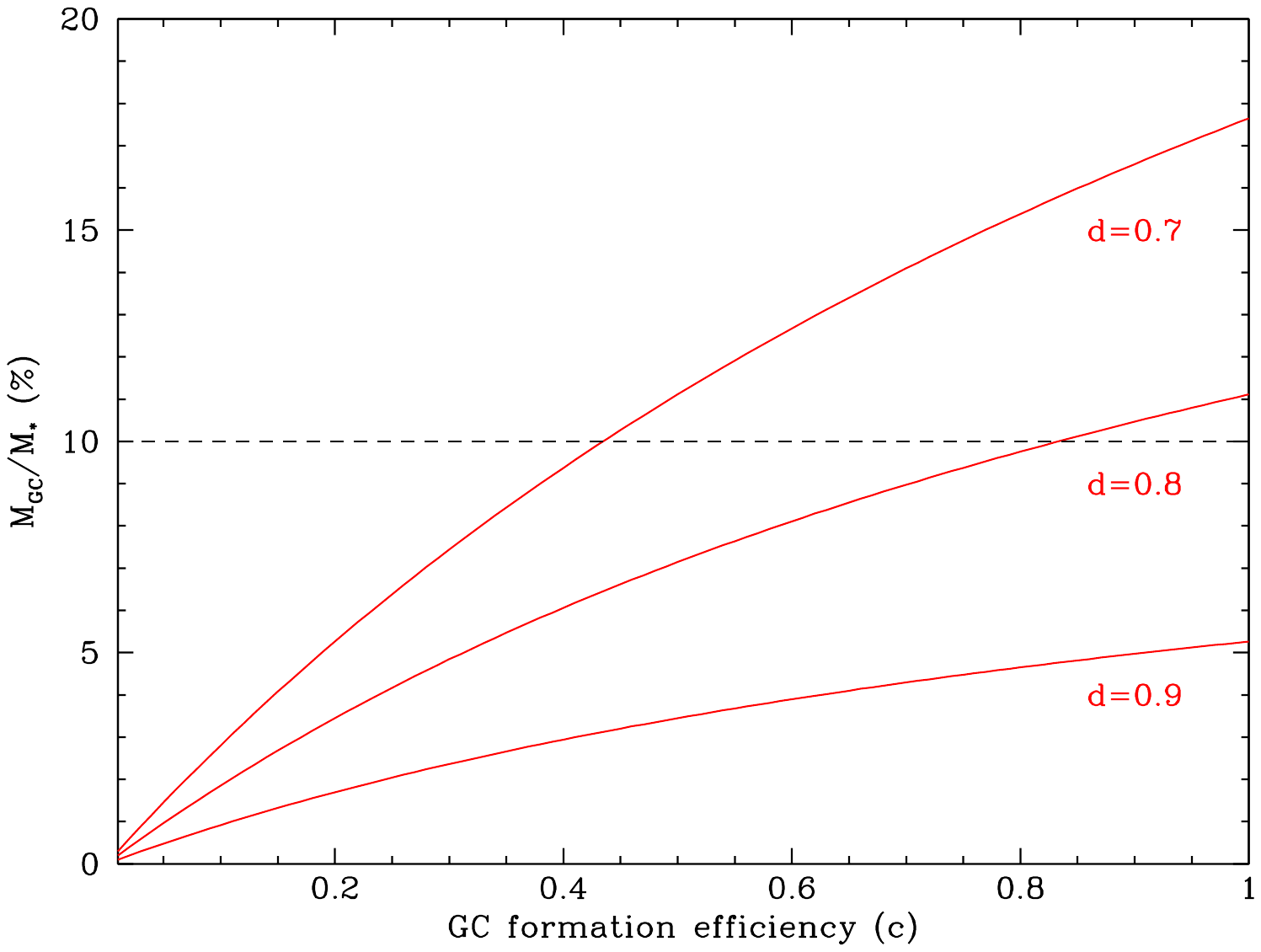}
	\caption{M$_{GC}$/M$_{\ast}$ as a function of GC formation efficiency ($c$).
 %, equivalent to 
 %M$_{GC}$/M$_{\ast}$ today vs 
 %M$_{GC}$/M$_{\ast}$ at formation. %A GC formation efficiency of one indicates equal fractions of the stellar mass in GCs and the host galaxy field stars.
 Three curves, from eq. 6,  show GC destruction fractions of $d$ = 0.7, 0.8 and 0.9. Destruction includes mass loss and tidal disruption. In order to reproduce the observed upper limit today for UDGs of 
 M$_{GC}$/M$_{\ast}$  $\sim$10\% (dashed line) a combination of either modest GC destruction rates ($\sim$70--80\%) and/or very high GC formation efficiencies ($\ge$40\%) are required. GC-poor galaxies with low 
 M$_{GC}$/M$_{\ast}$ ratios  suggest low GC formation efficiency and/or high destruction rates. 
  }
  \label{fig:one}
\end{figure}

\section{The Variation of Ultra Diffuse Galaxy Stellar Populations with Globular Cluster System Mass}

\begin{figure*}
	\includegraphics[width=0.8\linewidth,angle=-90]{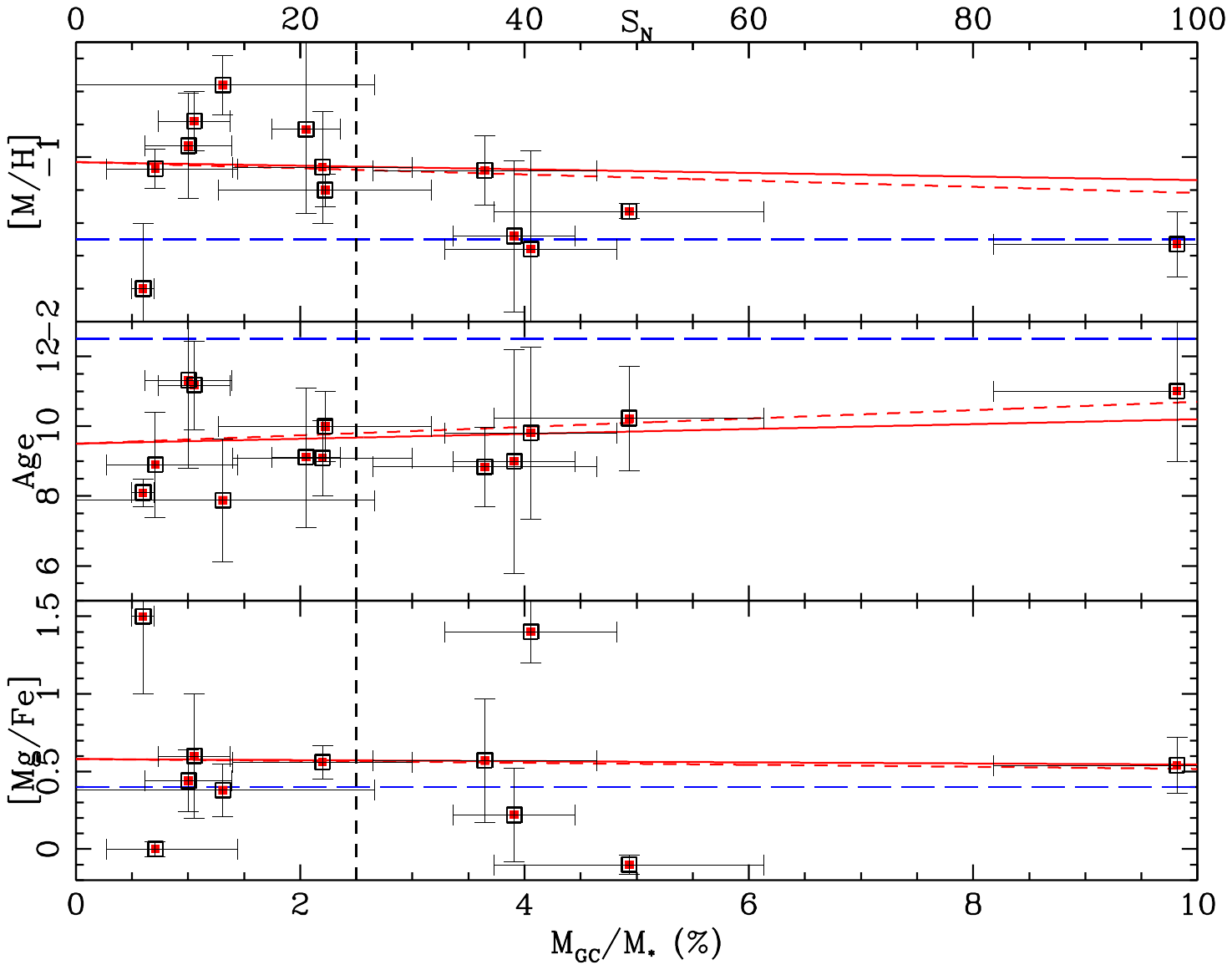}
	\caption{Galaxy stellar populations as a function of GC system mass to host galaxy stellar mass M$_{GC}$/M$_{\ast}$. Panels show total metallicity (in dex), age (in Gyr) and [Mg/Fe] (in dex) from top to bottom, respectively. 
 The red lines represent our simple model starting at M$_{GC}$/M$_{\ast}$ = 0 (puffy dwarfs) and as M$_{GC}$/M$_{\ast}$ increases the stellar population changes to be more GC-like with weights given by the fraction of disrupted GC mass (see text for details). The solid and dashed lines show the effect of increasing GC formation efficiency for a fixed the GC destruction fraction of d = 0.7 and 0.8 respectively. The long dashed blue lines indicate a constant GC-like stellar population 
(as might be the case if a UDG formed with GC-like stellar populations and had no further star formation). 
  Data points are our sample of UDGs (see Table \ref{tab:pars}), which are mostly located in high density environments, come from the catalogue of 
  Gannon et a. (2024). 
%\cite{2024MNRAS.531.1856G}. 
DGSAT~I is a field UDG with higher mass GCs than typically found 
Janssens et al. (2022);
%\citep{2022MNRAS.517..858J}; 
the value plotted of  
M$_{GC}$/M$_{\ast}$ = 0.6\% is a lower limit.
 The vertical line at 2.5\% represents a possible division between puffy dwarfs and failed galaxies 
 \citep{2016ApJ...822L..31P}. The top axis label shows approximate corresponding S$_N$ values. The data, and the models, show a decrease in metallicity and a slight increase in age with increasing M$_{GC}/$M$_{\ast}$ ratio. The [Mg/Fe] data shows considerable scatter with no clear trend.
}
\label{fig:two}
\end{figure*}

In order to study the stellar populations of UDGs we use the recent catalogue of 
\cite{2024MNRAS.531.1856G}. This compilation includes UDGs with spectroscopic stellar populations (largely from \citealt{2023MNRAS.526.4735F}) and/or velocity dispersions. It includes GC counts from a variety of sources (mostly derived from HST imaging). 
We include all UDGs from the catalogue with {\it both} stellar population and GC count information.
However, for NGC5846\_UDG1 we have decided to adopt the 
average of the four stellar population parameters from the literature. 
%follow  \cite{2022ApJ...927L..28D} who  determined a GC system of 54 $\pm$ 9 GCs and a stellar mass of 1.1 $\times$ 10$^8$ M$_{\odot}$ resulting in S$_M$ = 9.8\%. 
 %Thus we use this value rather than the 38 GCs derived from  single orbit HST imaging (Marleau **). 
 %For the integrated stellar population of its stars w
 These studies are \cite{2020A&A...640A.106M}, \cite{2023MNRAS.526.4735F}, \cite{2023A&A...676A..33H} and 
 \cite{2024MNRAS.529.3210B} (the first three are spectroscopic and the latter is from Spectral Energy Distribution fitting). 
 This gives an age of 11.0 $\pm$ 2.0 Gyr and metallicity [M/H] = --1.53 $\pm$ 0.2. For the alpha element ratio, the only value available is that of  \cite{2023MNRAS.526.4735F} who found [Mg/Fe] = 0.54 $\pm$ 0.18. We use the same GC count and stellar mass for UDG1 as listed in the 
\cite{2024MNRAS.531.1856G} catalogue, giving 
M$_{GC}$/M$_{\ast}$  = 9.8\%. 
We also add the recent GC counts from 
\cite{2024arXiv240907518J}
for PUDG\_R15 (13$\pm$5) and PUDG\_R84 (43$\pm$6). 
Unfortunately 
VCC 615 
(M$_{GC}$/M$_{\ast}$ = 8.3\%) and 
NGVSUDG-19 
(M$_{GC}$/M$_{\ast}$ = 5.4\%) are excluded as they lack stellar population information.

The stellar population values for each galaxy and its 
M$_{GC}$/M$_{\ast}$ ratio in our sample are summarised in Table \ref{tab:pars}.
Biases in the catalogue are discussed in \cite{2024MNRAS.531.1856G}. For example, other than DGSAT~1 
\citep{2022MNRAS.517..858J}  located in the field, the UDGs are located in high density environments. 
%groups and clusters. 
We also note that for DGSAT~I we assume the same mean GC mass as the other UDGs, although in this case there is evidence that some of the GCs may be more massive by a factor of $\sim$3x than is typical \citep{2022MNRAS.517..858J}. As noted in \cite{2024MNRAS.531.1856G}, the catalogue has a bias against GC-poor UDGs.  We note that metallicity, in particular, has a strong dependence on stellar mass for dwarf galaxies (e.g. \cite{2019ARA&A..57..375S}) but 
our UDG sample has a relatively small range in mass and we do not attempt to correct for the mass-metallicity relation. 
While ages and metallicities are available for more UDGs from SED fitting studies, such studies lack [Mg/Fe] measurements and tend to have large uncertainties on their 
M$_{GC}$/M$_{\ast}$ ratios.

Fig. \ref{fig:two} shows the observed stellar populations for our UDG sample as a function of their observed 
M$_{GC}$/M$_{\ast}$ ratio, ranging from 
M$_{GC}$/M$_{\ast}$ = 0\% (puffy dwarf) to 10\% (failed galaxy). The UDG with the highest ratio of 9.8\% is NGC5846\_UDG1 (see Table 1). 
The panels show metallicity [M/H], age (in Gyr) and [Mg/Fe]. %Two UDGs (DGSAT~I and Yagi358) with [Mg/Fe] $\sim$ 1.5 
%(\citealt{Martin2019}; \citealt{FerreMateu2018})
%are not shown for clarity purposes in the [Mg/Fe] %panel. 
The data show a decrease in metallicity ($\sim$0.45 $\pm$ 0.1 dex) and a weak hint of older ages  ($\sim$0.6 $\pm$ 1 Gyr) as the ratio of GC system mass to stellar mass increases.
%Excluding NGC5846\_UDG1 removes any age trend but the metallicity decrease remains. 
There is no clear trend for [Mg/Fe], which shows a large scatter in values. It is not clear if this large scatter in [Mg/Fe] is due to observational uncertainty or represents an intrinsic variation in UDGs. 

As well as examining how the stellar population properties of UDGs vary with GC system mass, we also wish to compare the observations with predictions from our simple model of GC formation and destruction.
The stellar populations of a puffy dwarf UDG (i.e. a dwarf galaxy that is puffed up in size and made more diffuse with a low M$_{GC}$/M$_{\ast}$ ratio) 
might be expected to resemble that of a classical dwarf galaxy if the puffing-up process does not modify its stellar population. This is probably true when puffy dwarfs are simply the high spin tail of a dwarf galaxy distribution \citep{2016MNRAS.459L..51A} 
but less so if multiple supernovae bursts \citep{2017MNRAS.466L...1D} is the formation mechanism. 
Failed galaxies (with high
M$_{GC}$/M$_{\ast}$ ratios) 
might be expected to have stellar populations similar to those of metal-poor GCs if their GCs formed very efficiently and were also largely destroyed so that GC stars now contribute to the overall galaxy stellar population
\citep{FerreMateu2018}. 
Similarly, mass loss at the proto-GC stage would also contribute GC stars to the field star stellar population.

In Fig. \ref{fig:two} we also show our simple model as a function of 
M$_{GC}$/M$_{\ast}$, which is given by eq. 6. We start by assigning a stellar population value for 
UDGs with M$_{GC}$/M$_{\ast}$ = 0. Rather than use a proxy such as classical dwarf galaxies, we use the average value for the 5 sample UDGs with 
M$_{GC}$/M$_{\ast}$ $<$ 1.5\% to represent a puffy dwarf. This can be improved in the future when more spectroscopic studies of GC-poor UDGs are conducted. From these 5 UDGs we find mean values of [M/H] = --1.03, age = 9.5 Gyr and [Mg/Fe] = 0.58 dex. We use these values as the `zero point' for the stellar population  of a UDG with M$_{GC}$/M$_{\ast}$ = 0, i.e. a puffy dwarf. 

As GCs are disrupted they contribute to the integrated stellar populations of the UDG. 
To characterise the mean stellar population of old, metal-poor GCs we assume a mean age of 12.5$\pm$1 Gyr, mean metallicities [M/H] of --1.5$\pm$0.3 and [Mg/Fe] of +0.4$\pm$0.1  from the compilation of Milky Way GCs by \cite{2018A&A...620A.194R}.
To account for the contribution of disrupted GCs to the stellar population of the host galaxy
%To mimic that disruption process,  
%M$_{GC}$/M$_{\ast}$ increases (via e.q 6) 
we weight the stellar population using Eq. 7. For example, if the destruction fraction $d$ = 0.7 and the formation efficiency $c$ = 0.4, then the resulting stellar population has a contribution from disrupted GCs (age 12.5 Gyr) weighted by 0.22 (e.q 7) and the original stellar population of the stars  
(age 9.5 Gyr) weighted by 1--0.22 = 0.78 to get 0.22x12.5 + 0.78x9.5 = 10.2 Gyr. Following Fig. \ref{fig:one}, we show two model tracks in Fig. \ref{fig:two} for a fixed destruction fraction of $d$ = 0.7 (solid line) and $d$ = 0.8 (dashed line) with the GC formation efficiency ($c$) allowed to vary.
We remind the reader that destruction in the 
\cite{2024MNRAS.527.2765M}
models includes GC mass loss.
The model curves reveal a decrease in metallicity and increase in age,  
approaching a GC-like stellar population (and the location of NGC5846\_UDG1) at the highest M$_{GC}$/M$_{\ast}$ ratio. The effect of increasing 
destruction from $d$ = 0.7 to 0.8 further decreases the metallicity and increases the age of the model tracks. The model predicts a relatively small change in [Mg/Fe] ratio with increasing M$_{GC}$/M$_{\ast}$ ratio.

An additional line is shown in Fig. \ref{fig:two} which represents a constant GC-like stellar population. If the initial stars of a UDG formed from the same metal-poor gas at early times as the GCs and is immediately quenched (i.e. no ongoing star formation) then one would expect GC-like stellar populations. 
Further contributions from GC mass loss and/or  disrupted GCs would then not change the integrated stellar population. Thus, this situation represents a limit to the expected stellar populations. 
The UDG metallicity and age data tend to be more metal-rich (by $\sim$0.3 dex) and younger (by $\sim$1.5 Gyr) than the pure GC stellar population, as shown by the long dashed line in Fig. \ref{fig:two}. If supported by additional data, this would suggest that some star formation occurred after the main epoch of GC formation even for failed galaxies.

More data are needed at M$_{GC}$/M$_{\ast}$ high ratios
to test the model trends with increasing M$_{GC}$/M$_{\ast}$ ratio. More data at very low ratios
would allow for a better `zero point' for the stellar populations of puffy dwarfs. 
More observational data might help to 
determine whether a ratio of 2.5\%,  as advocated by 
\cite{2016ApJ...822L..31P},  represents a real division between puffy dwarfs and failed galaxies. Although we note that our simple model is continuous and does not predict a clear distinction. 
Another interesting approach is to compare the mean colours of GCs with that of the galaxy stars as done recently by 
\cite{2024arXiv240907518J} for a sample of Perseus UDGs.

{\it Within the limitations and assumptions of our simple model, we suggest that GC-rich UDGs with high 
M$_{GC}$/M$_{\ast}$ ratios (failed galaxies) are likely the result of very high GC formation efficiencies with modest rates of GC destruction. Additional data are needed to further test and refine the simple model. }

\begin{table*}
  \centering
 \begin{tabular}{l|c|c|c|c}
  \hline
 Name & M$_{GC}$/M$_{\ast}$ (\%) & [M/H] (dex) & Age (Gyr) & [Mg/Fe] (dex)\\ 
 \hline
DF44 & 4.9          (1.2)  &     -1.33        (+0.05,         -0.04)    &   10.23         (1.50) &        -0.1        (0.06)\\
DF07 & 1.1       (0.3)   &    -0.78        (0.18)    &   11.18        (1.27)  &        0.6         (0.4)\\
DF17 & 2.1       (0.3)    &   -0.83        (+0.56,         -0.51)     &   9.11           (2.0)   &     --        (--)\\
DF26 & 1.3        (1.3)     &  -0.56        (0.18)     &   7.88        (1.76)    &    0.38        (0.17)\\
DFX1 & 3.6            (1.0)     &  -1.08        (0.21)    &    8.84        (1.13)    &    0.57         (0.4)\\
DGSAT~I & 0.6          (0.1)$^{\dagger}$   &     -1.8         (0.4)   &      8.1         (0.4)   &      1.5         (0.5)\\
Hydra~I-UDG11 & 2.2       (0.9)    &     -1.2         (0.1)      &    10           (1.0)    &    -- (--) \\
NGC1052-DF2 & 0.7        (+0.7,       -0.4)     &  -1.07        (0.12)     &    8.9         (1.5)   &        0        (0.05)\\
NGC5846\_UDG1 & 9.8        (1.6)     &  -1.53         (0.2)     &     11           (2.0)    &    0.54        (0.18)\\
%5.4        2.419       2.419        -999         -99         -99        -999         -99         -99        -999         -99         -99
%17.3846        13.23       13.23        -999         -99         -99        -999         -99         -99        -999         -99         -99
PUDG-R15 & 1.0      (0.4)   &    -0.93        (0.32)   &     11.32        (2.5)   &     0.44         (0.2)\\
PUDG-R84 & 3.9       (0.6)    &   -1.48        (0.46)   &    8.99         (3.2)     &   0.22         (0.3)\\
%0.9       0.6687      0.6687        -999         -99         -99        -999         -99         -99        -999         -99         -99
%1.7        1.106       1.106        -999         -99         -99        -999         -99         -99        -999         -99         -99
VCC1287 & 2.2          (0.8)    &   -1.06        (0.34)   &     9.09        (1.07)    &    0.56        (0.11)\\
%8.3         2.63        2.63        -999         -99         -99        -999         -99         -99        -999         -99         -99
%4.3        2.301       2.301        -999         -99         -99        -999         -99         -99        -999         -99         -99
%23.7273            9           9        -999         -99         -99        -999         -99         -99        -999         -99         -99
%4.5        2.379       2.379        -999         -99         -99        -999         -99         -99        -999         -99         -99
Yagi358 & 4.1       (0.8)    &   -1.56         (0.6)    &    9.81        (2.46)    &     1.4         (0.2)\\
   \hline
 \end{tabular}
  \caption{Globular cluster system and stellar population properties for UDGs. Column 1: UDG name, column 2: GC system mass to galaxy stellar mass, column 3: galaxy total metallicity, column 4: galaxy age, column 5: galaxy [Mg/Fe] ratio. Uncertainties are given in round brackets. $^{\dagger}$ 
  Janssens et al. (2022) 
  %\cite{2022MNRAS.517..858J} 
  found DGSAT~I to have higher mass GCs than the typical mean value assumed here of 2 $\times$ 10$^5$ M$_{\odot}$.     
  }
  \label{tab:pars}
\end{table*}

%      & [M/H] & Age & [Mg/Fe] & log M$_{\ast}$\\
%      & (dex) & (Gyr) & (dex) & (M$_{\odot}$) \\
%     \hline 
%Classical dwarfs & --0.48$\pm$0.35 & 7.6$\pm$3.9 & 0.24$\pm$0.32 & 8.8$\pm$0.6\\
%Classical dwarfs & --0.3$\pm$0.3 & 2$\pm$1 & 0.0$\pm$0.1 & 9.0\\
%Metal-poor GCs & --1.5$\pm$0.3 & 12.5$\pm$1.0 & 0.4$\pm$0.1 & 5.3\\
%\hline
%Anna's GCs & --1.27$\pm$0.51 & 12.1$\pm$1.1 & 0.31$\pm$0.13 & 5.8$\pm$0.6\\
%Forbes2020 dEs & -0.3 & 4 & 0.0 & --\\
%Forbes2020 GCs & -2.2 & 10 & 0.8 & --\\
%MPGCs & -2 & 13 & 0.4 & 5.3\\
%Sansom08 & -1.3 to 0.4 & $<$3 & -0.03 to 0.075 & $\sim$9\\
%NUDGEs & --0.9$\pm$0.2 & 9.0$\pm$1.3 & -- & 7.5$\pm$0.3\\
%\hline
%Anna's UDGs & --1.0$\pm$0.4 & 8.3$\pm$3.3 & 0.51$\pm$0.32 & 8.4$\pm$0.3\\
%MatlasUDGs & --1.2$\pm$0.2 & 7.1$\pm$1.8 & 
%-- & 7.6$\pm$0.3\\
%Buzzo22UDGs & --1.2$\pm$0.2 & 8.2$\pm$1.4 & --& 8.1$\pm$0.3\\
%    \hline
%  \end{tabular}
%  \caption{Mean stellar population properties for our reference samples of classical dwarfs and metal-poor globular clusters. Lower portion of table gives the mean propertie of our UDG sample from \cite{2024MNRAS.531.1856G}.
  %and Buzzo25 (from SED fitting). 
%    }
%  \label{tab:fitparams}
%\end{table}

\section{Possible Examples of Failed Galaxies}
%\subsection{NGC5846\_UDG1: A Prototype Failed Galaxy?}

%The defining characteristic of a failed galaxy is that it resides in a massive halo but for some reason formed few stars, giving it a high M$_{halo}$/M$_{\ast}$ ratio so it lies off the standard stellar mass-halo mass relation with an overly massive halo. Given the high GC formation efficiencies seen in recent JWST observations and modest destruction rates, high M$_{GC}$/M$_{\ast}$ (= S$_M$) are expected. If there was little, or no, star formation over cosmic time, then the integrated stellar population of a failed galaxy would resemble that of a metal-poor GC (old ages, low metallicity, high [Mg/Fe]) and uniform blue colours. The host galaxies  would tend to have low stellar metallicities for their stellar mass.   

Perhaps the best candidate for a failed galaxy is 
NGC5846\_UDG1. It has a rich GC system of 54$\pm$9 GCs 
(M$_{GC}$/M$_{\ast}$ = 9.8\%), which 
according to the GC number -- halo mass relation 
\citep{Burkert2020}, 
%or a cored halo profile from measured dynamics \citep{2024MNRAS.528..608F}, 
implies a large M$_{halo}$/M$_{\ast}$ ratio  
%NGC5846\_UDG1 
placing it off the standard stellar mass - halo relation. 
Its GC system extent is comparable to that of the host galaxy, ie R$_{GC}$/R$_e$ = 0.8$\pm$0.2 \citep{2022ApJ...927L..28D}. 
Using MUSE on the VLT, \cite{2020A&A...640A.106M} found consistent (old) ages and (low) metallicities for the GCs and stars. This suggests that the stars of the galaxy and those in the GCs formed at a similar time from the same enriched gas. This would be as expected if the stars of disrupted GCs now contribute significantly to the stellar field star population. 
\cite{2020A&A...640A.106M} did not measure [Mg/Fe] ratios but \cite{2023MNRAS.526.4735F} found super-solar [Mg/Fe] for the stars, which is again consistent with those of metal-poor GCs. It is a nearly round galaxy (b/a = 0.9) with no evidence of rotation
\citep{2021MNRAS.500.1279F}. 
All of these properties match those expected of a failed galaxy, as listed in section 2.1. 

As well as VCC 615 and NGVSUDG-19 mentioned above, 
\cite{2023MNRAS.526.4735F} have noted four UDGs with low stellar metallicities for their stellar mass and hence potential failed galaxy candidates. They are: DGSAT~I (field), DF44 (Coma), Yagi358 (Coma) and PUDG-R84 (Perseus). They also all have old ages. Their 
M$_{GC}$/M$_{\ast}$ ratios are: 0.6\%, 4.9\%, 4.1\% and 3.9\% (see Table 1). DGSAT~I reveals an extremely high [Mg/Fe] ratio and a very high dynamical to stellar mass ratio
\citep{2019MNRAS.484.3425M}. 
In the case of DF44, a total halo mass measurement is available from its radial kinematics 
\citep{2019ApJ...880...91V} 
which indicates an overly massive halo (large M$_{halo}$/M$_{\odot}$ ratio). These galaxies are potential candidates for failed galaxies and warrant further study. 

There may also be some parallels between failed galaxy UDGs and much lower stellar mass dwarfs. For example, the 
Local Group dwarf Eridanus II has an  
M$_{GC}$/M$_{\ast}$ ratio of $\sim$4\% today from its only GC. 
\cite{2023ApJ...948...50W} suggest that at birth this single GC, before mass loss, represented $\sim$10\% of the galaxy stellar mass. The presence of other GCs would make this mass fraction at formation even higher, and contribute to the field star population when disrupted. They found that the field stars of Eridanus II have  old ages (13.5$\pm$0.3  Gyr) and very low metallicity ([Fe/H] $\sim$ --2.6$\pm$0.15). They also found the GC to have the same age and metallicity within uncertainties. After initial star formation at an early epoch, the galaxy was effectively quenched. 
%by reionisation between redshifts 10 to 6. 
\cite{2018MNRAS.481.5592F} derived a high halo mass to stellar mass ratio for Eridanus II, similar to those inferred for UDG failed galaxy candidates.

%\section{Discussion}

\section{Summary and Conclusions}

Observations of UDGs reveal that some of them have many more GCs per unit starlight than classical dwarf galaxies of the same stellar mass. These high M$_{GC}$/M$_{\ast}$ UDGs have been dubbed failed galaxies, whose key property is an inferred overly massive halo for their stellar mass (i.e. high M$_{halo}$/M$_{\odot}$ ratio). 

Simulations indicate that it is high gas densities and/or gas accretion rates that lead to higher fractions of mass in bound star clusters relative to field stars. Such conditions are common at early epochs. From observations, it is difficult to infer the natal gas pressure in an infant galaxy or to track its accretion rate, but recent observations with JWST of high redshift lensed galaxies have revealed very high ratios (40--70\%) of %M$_{GC}$/M$_{\odot}$
the mass in bound, high density star clusters relative to that of the stellar mass of the host galaxy. These appear to be GCs forming up to 13.5 Gyr ago in, and around, a host galaxy with a high M$_{GC}$/M$_{\ast}$ ratio. 

To understand GC destruction over cosmic time we must rely on simulations. 
The recent work of 
\cite{2024MNRAS.527.2765M} modelled GC mass loss and tidal disruption in dwarf galaxies. They found lower GC destruction (higher survival fractions at the present day) in lower mass dwarfs and for dwarfs of lower surface density. This may help explain why UDGs, of low surface density, 
have higher 
M$_{GC}$/M$_{\ast}$ ratios today than classical (higher surface density) dwarfs.

Using these constraints on 
GC formation efficiency 
and destruction fractions 
%found by the 
%\cite{2024MNRAS.527.2765M} simulations, 
we have created a simple model to help understand the stellar populations from GC-poor puffy dwarfs and GC-rich failed galaxies.  
%whether GC-rich UDGs had experienced high 
%GC formation and/or modest GC destruction. 
Our simple GC formation/destruction model has different expectations for the stellar populations of the host galaxy, assuming that disrupted GCs contribute to the field stars of the galaxy and there  is no ongoing star formation in the galaxy (i.e. it quenched early). We found that failed galaxy UDGs, with high 
M$_{GC}$/M$_{\ast}$ ratios, are likely the result of very high GC formation efficiencies combined with modest rates of GC destruction. Ultimately, full hydrodynamical simulations of UDGs in a cosmological context are required that incorporate high rates 
of GC formation, early quenching of star formation and GC destruction over cosmic time.  
Meanwhile, our simple model can be refined and 
tested as more UDG stellar population data becomes available.

\section*{Acknowledgements}

We wish to thank the anonymous referee for their comments that allowed us to improve the paper.
We thank other members of the AGATE team for their help and useful discussions: W. Couch, L. Haacke, J. Pfeffer. 
%A. Romanowsky, S. Danieli, L. Buzzo, A. Ferre-Mateu, L. Haacke, J. Pfeffer, J. Brodie, O. Gnedin, R. Remus, S. Kim, J. Read and M. Collins for useful discussions.
We thank R. Remus, L. Kimmig, J. Read for helpful discussions. 
DF and JB thank the ARC for support via DP220101863 and DP200102574.
AFM has received support from RYC2021-031099-I and PID2021-123313NA-I00 of 
MICIN/AEI/10.13039/501100011033/FEDER, UE, NextGenerationEU/PRT. 
MLMC acknowledges support from STFC grants ST/Y002857/1 and ST/Y002865/1.
AJR was supported by National Science Foundation grant AST-2308390.

\section*{Data Availability}

This study made use of publicly available data in the literature. In particular, sources collected by 
\cite{2024MNRAS.531.1856G}: \cite{mcconnachie2012, 
2015ApJ...798L..45V, 
%vanDokkum2015, 
Beasley2016, Martin2016, Yagi2016, MartinezDelgado2016, vanDokkum2016, 
2017ApJ...844L..11V, 
%vanDokkum2017, 
Karachentsev2017, vanDokkum2018,  Toloba2018, Gu2018, 
2018ApJ...862...82L, 
%Lim2018, 
RuizLara2018, Alabi2018, FerreMateu2018, 
2018MNRAS.481.5592F, 
%Forbes2018, 
2019MNRAS.484.3425M, 
%MartinNavarro2019, 
Chilingarian2019, Fensch2019, Danieli2019, 
2019ApJ...880...91V, 
%vanDokkum2019b, 
torrealba2019, Iodice2020, Collins2020, 
2020A&A...640A.106M, 
%Muller2020, 
Gannon2020, 
2020ApJ...899...69L,
%Lim2020, 
Muller2021, 
2021MNRAS.500.1279F, 
%Forbes2021, 
Shen2021, Ji2021, Huang2021, 
2021MNRAS.502.3144G, 
%Gannon2021, 
Gannon2022, Mihos2022, 
2022ApJ...927L..28D, 
%Danieli2022, 
Villaume2022, Webb2022, 
2022MNRAS.511.4633S, 
%Saifollahi2022, 
2022MNRAS.517..858J, 
%Janssens2022, 
Gannon2023, 
2023MNRAS.526.4735F, 
%FerreMateu2023, 
2023ApJ...951...77T, 
%Toloba2023, 
Iodice2023, Shen2023}

%Reasonable request for data can be made to the first author.

%\software{Astropy \citep{Astropy2018}, Photutils \citep{Bradley2021}, numpy \citep{Numpy2011}, scipy \citep{Virtanen2021}, matplotlib \citep{Hunter2007}.}

%\facilities{\textit{HST} (ACS)}

\bibliographystyle{mnras}
\bibliography{failed}{}

\bsp	% typesetting comment
\label{lastpage}
\end{document}